\documentclass[12pt]{spieman}  

\usepackage{amsmath,amsfonts,amssymb}
\usepackage{graphicx}
\usepackage{setspace}
\usepackage{tocloft}
\usepackage{subcaption}
\usepackage{comment}

\title{Astronomical Image Processing Benchmark Study for Various Telescope Aperture Shapes}

\author[1,*]{Jyotika Roychowdhury}
\author[2]{Kevin Derby}
\author[2,3]{Daewook Kim}
\affil[1]{Department of Physics, Indian Institute of Technology Delhi, New Delhi 110016, India}
\affil[2]{Wyant College of Optical Sciences, University of Arizona, 1630 E University Blvd, Tucson, AZ 85721, USA}
\affil[3]{Department of Astronomy and Steward Observatory, University of Arizona, 933 N Cherry Ave, Tucson, AZ 85721, USA}

\cftpagenumbersoff{figure}
\cftpagenumbersoff{table} 
\begin{document} 
\maketitle

\begin{abstract}
We explore the impact of different telescope apertures on the image simulation and deconvolution processes within the context of a synthetic star field. Using HCIPy and Python programming, we modelled six telescope apertures namely Circular, Hexagonal, Elliptical (with horizontal and vertical major axes), segmented hexagonal (JWST), and obstructed circular (HST). We calculated Point Spread Functions (PSFs) for each aperture, incorporating surface shape-induced wavefront aberrations, convolved them with a synthetic star field spanning a range of brightness magnitudes, and introduced photon and detector noise layers to simulate realistic imaging conditions. Subsequent deconvolution using the Richardson-Lucy algorithm allowed for an analysis of deconvolution accuracy based on parameters like average distance between stars and differences in the number of stars between original and deconvolved images. Results indicate that the choice of telescope aperture significantly influences both simulated images and deconvolution outcomes, with brightness magnitude also playing a crucial role. The study highlights the necessity of optimizing image processing pipelines and Deconvolution algorithms tailored to each aperture shapes and their corresponding PSFs, emphasizing the pivotal role of aperture selection and optimization in achieving accurate astronomical imaging performance. 
\end{abstract}

\keywords{Telescope Apertures, Wavefront sensing, High Contrast Imaging, PSF Convolution and Deconvolution, Image Processing}

{\noindent \footnotesize\textbf{*}Jyotika Roychowdhury,  \linkable{ph1200702@iitd.ac.in} }

\begin{spacing}{1}   

\section{Introduction}
\label{sect:intro}  
Our understanding of the universe is based on astronomical observations, which are shaped by the telescopes used to make these observations. The telescope aperture thus serves as the primary element which gathers information in the form of light and plays a crucial role in the entire process. As telescope technology advances - from the groundbreaking James Webb Space Telescope (JWST) to the novel Nancy Grace Roman Space Telescope - there is a need to find optimal telescope apertures and develop computational methods to efficiently process its data. Difference in aperture designs has a significant effect on the astronomical images that are produced. Understanding how these aperture configurations impact image quality and the efficiency of post-processing algorithms is therefore very important for advancing observational astronomy and refining image processing techniques.

\subsection{Precision Astrometry}

Precision astrometry is used for measuring the positions, motions, and distances of celestial objects. The accuracy of astrometric data is important for studying stellar dynamics, understanding the structure of galaxies, and discovering exoplanets through astrometric techniques. However, challenges arise as the PSF of different apertures introduce complexities in achieving high precision. Astrometry algorithms usually use a reference PSF for the entire field of view. Different apertures can lead to different field-dependent aberrations, which in turn can cause variations in the PSF. Due to these variations, the reference PSF is only accurate for a small portion of the image called the isoplanatic patch, where the PSF remains relatively consistent.
The choice of aperture thus influences the PSF and affects the accuracy of astrometric measurements. Therefore, finding the optimal aperture becomes crucial for precision astrometry.

\subsection{Improvement of Processing Pipelines}

The development of advanced computational algorithms and post-processing pipelines amplifies the imaging capabilities of existing hardware. It also stands as a cost-effective strategy for maximizing the potential of observational data. Through sophisticated image reconstruction techniques, these algorithms not only enhance the resolution and accuracy of astronomical images but also contribute to extracting richer information without the need for substantial hardware upgrades. Thus optimizing data from telescope apertures and making computational advancements is a prudent investment in extending the lifespan and efficiency of current telescopic systems while facilitating groundbreaking discoveries in astronomy.

\section{Methodology}

In this section, we present our methodology for the simulation of the pipeline, which consists of various steps as shown in the flowchart below - 

\begin{figure}[htp]
   \begin{center}
        \includegraphics[scale=0.5]{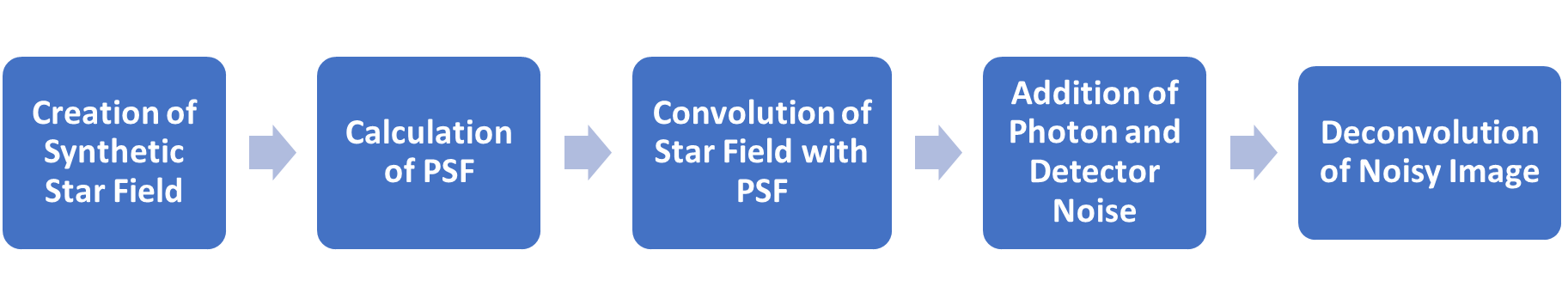}
   \end{center}
   \caption{Methodology of the astronomical image processing Pipeline for the benchmark study}
\end{figure}

\subsection{Creation of Synthetic Star Field Image}
\label{sect:title}
We generate synthetic star fields to have a testing ground for investigating the effects of different telescope apertures on image deconvolution processes. The two main features of our star fields are: 
\begin{itemize}
    \item \textbf{Gaussian distribution of positions of stars:} Our synthetic star field follows a Gaussian distribution of positions of stars. The positions of stars are more concentrated toward the center of the field and gradually decrease in density as we move away from the center. One of the main reasons to choose this configuration is to make our simulations similar to the structure of a globular cluster, thereby making it analogous to the real world observations. Following a Gaussian configuration also led to the stars being away from edges, which helped us to avoid complications during PSF convolution and post-processing.  
    \item \textbf{Uniform brightness:} The magnitudes of brightness of all the stars in the synthetic star field are assumed to be equal in this benchmark study. Across different simulations, the magnitude of the star field is varied within a range of 0 to 14. The same condition is applied to all simulation cases for the fair comparison studies.
\end{itemize}

\begin{figure}[htp]
   \begin{center}
        \includegraphics[scale=0.5]{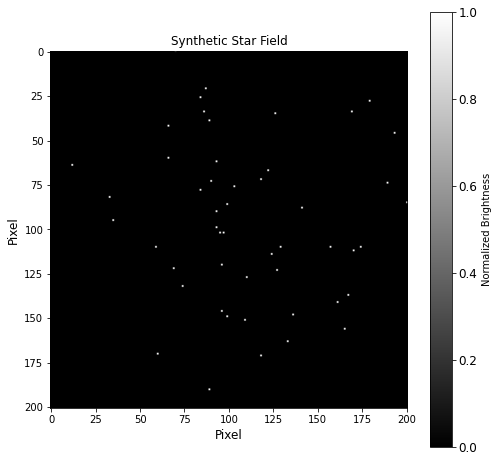}
   \end{center}
   \caption{Synthetic Star Field Image}
\end{figure}

\subsection{Calculation of the realistic PSF}
\label{sect:title}
In this section, we discuss our process for modelling six space-based telescopes - five 6.5 m diameter telescopes with circular, hexagonal, elliptical with a horizontal major axis, elliptical with a vertical major axis and hexagonal segmented apertures (similar to JWST) and one 2.4 m diameter space-based telescope with an obstructed circular aperture (similar to HST). We set the effective focal length of the telescope to 70 m and the wavelength of light ($\lambda$) to 1000 nm. We assume to sample each \(\frac{\lambda}{D}\) with 2 pixels.
These six telescope apertures are created in Python using methods of physical optics propagation. High Contrast Imaging in Python \cite{hcipy-docs,inproceedings} (HCIPy) is an
open-source Python library that performs physical optics modeling using either Fraunhofer or
Fresnel approximations of scalar diffraction. Using HCIPy, we create a Wavefront object based on the telescope pupil and wavelength. We then calculate the total flux of the wavefront using the formula-
\begin{center}
    \(F = B \times A \times QE \times T \times Z \times 10^\frac{-M}{2.5}\)
\end{center}
In the above equation, \(F\) represents the Total Flux. The Spectral Band, denoted by \(B\), is calculated as the product of the wavelength (\(\lambda\)) and the Bandwidth. In our case, we have taken a bandwidth of 20\%.
\[ B = \lambda \times \text{{Bandwidth}} \]
\(A\) corresponds to the total collecting area of the telescope aperture. The quantum efficiency of the detector, denoted as \(QE\), is set at 90\%. \(T\) represents the throughput or efficiency of the aperture, which measures the amount of light collected by the telescope that actually reaches the detector. We assume this to be 20\%.
\(M\) represents the magnitude of the starfield, which is a measure of its brightness. Magnitude operates on an inverse scale, meaning that lower values indicate brighter stars. \(Z\) denotes the Vega Flux Zero Point \cite{martiniusefuldata} and it has a value of \(702\) photons \(\text{cm}^{-2} \text{s}^{-1} \text{A}^{-1}\). The Vega Flux Zero Point serves as a reference point for calibrating astronomical measurements, providing a standard for the observed flux from the star Vega. \\
We also account for surface aberrations in the telescope aperture, taking the Peak-To-Valley length to be equal to \(\frac{\lambda}{4}\). We then create a Fraunhofer Propagator and pass our wavefront through this propagator to get the focal image. The intensity distribution in the focal plane gives us the PSF.

\begin{figure}[htp]
        \centering
        \includegraphics[scale=0.3]{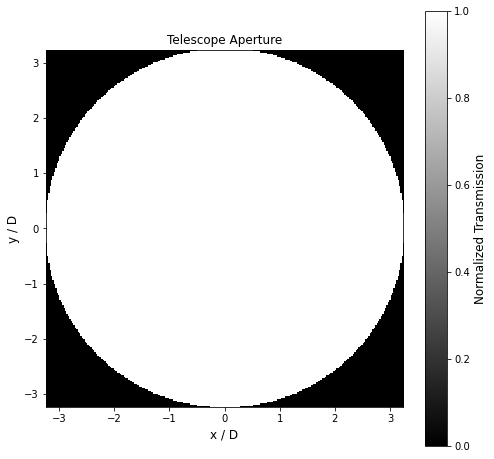}
        \includegraphics[scale=0.3]{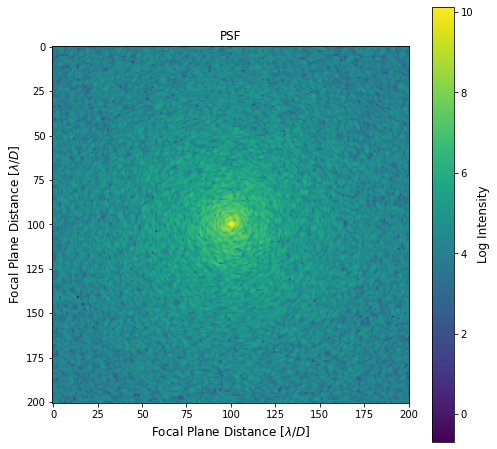}
   \caption{Circular Aperture and its PSF}
\end{figure}

\begin{figure}[htp]
        \centering
        \includegraphics[scale=0.3]{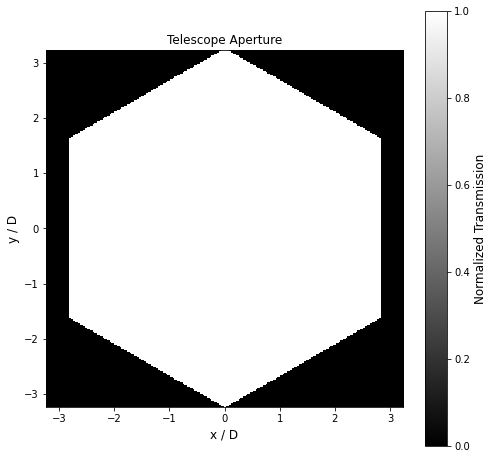}
        \includegraphics[scale=0.3]{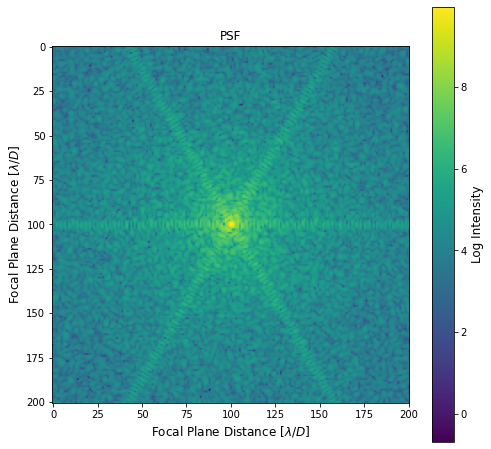}
   \caption{Hexagonal Aperture and its PSF}
\end{figure}

\begin{figure}[htp]
        \centering
        \includegraphics[scale=0.3]{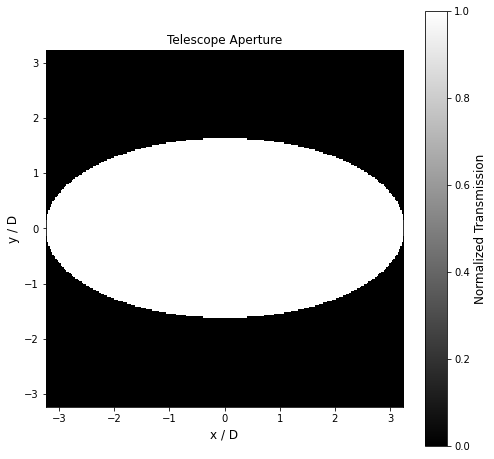}
        \includegraphics[scale=0.3]{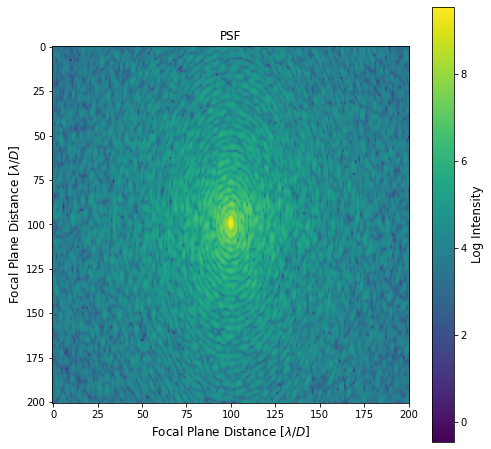}
   \caption{Elliptical Aperture with horizontal major axis and its PSF}
\end{figure}

\begin{figure}[htp]
        \centering
        \includegraphics[scale=0.3]{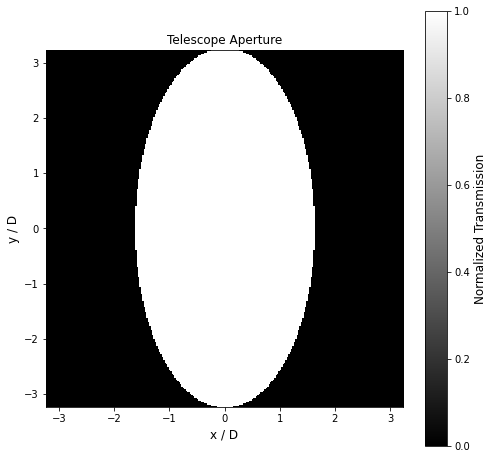}
        \includegraphics[scale=0.3]{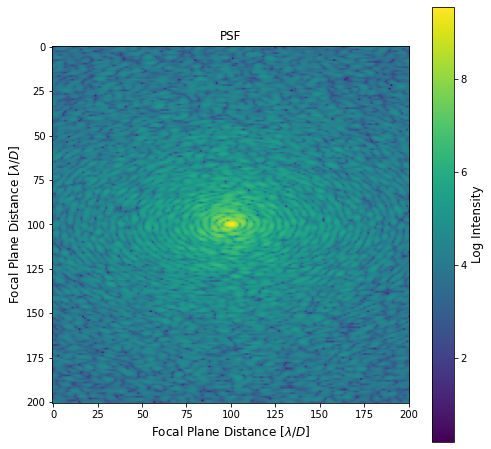}
   \caption{Elliptical Aperture with vertical major axis and its PSF}
\end{figure}

\begin{figure}[htp]
        \centering
        \includegraphics[scale=0.3]{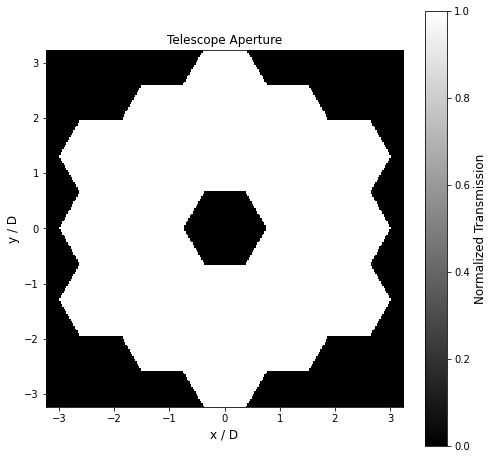}
        \includegraphics[scale=0.3]{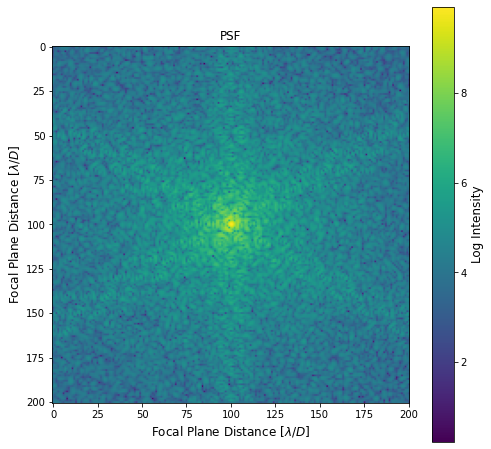}
   \caption{JWST Aperture and its PSF}
\end{figure}

\begin{figure}[htp]
        \centering
        \includegraphics[scale=0.3]{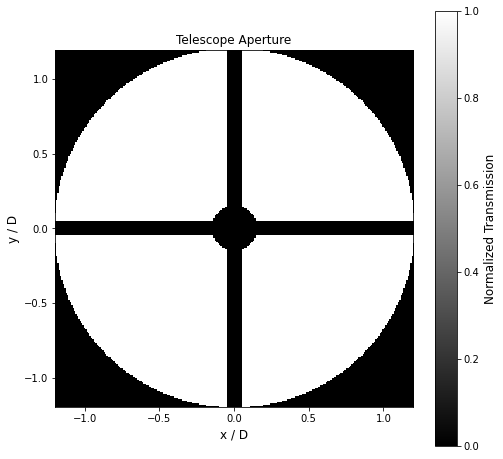}
        \includegraphics[scale=0.3]{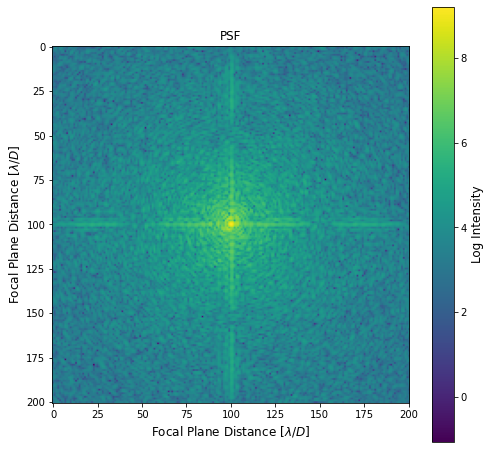}
   \caption{HST Aperture and its PSF}
\end{figure}

\newpage
\subsection{Convolution of Synthetic Star Field}
\label{sect:title}
The next step involves convolving the star field with the obtained PSF. We can notice distinct small patterns corresponding to the used aperture within the convolved star field.

\begin{figure}[htp]
\begin{tabular}{c}
        \includegraphics[scale=0.3]{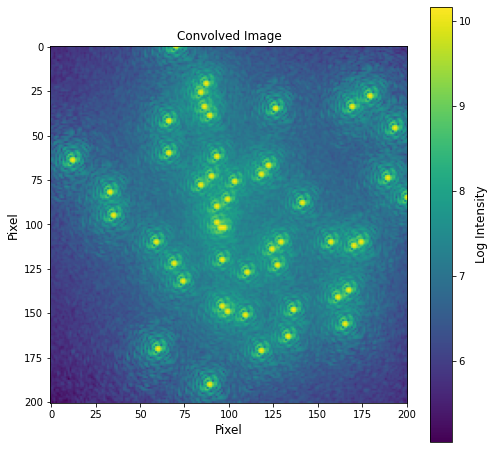}
        \includegraphics[scale=0.3]{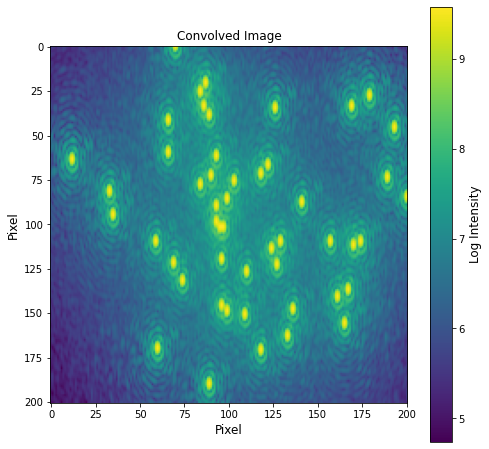}
        \includegraphics[scale=0.3]{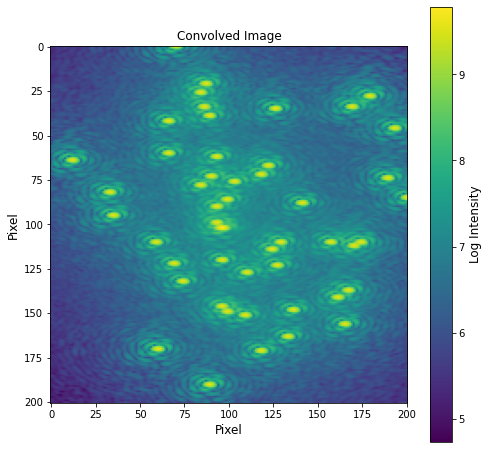}\\
(a) \hspace{4.5 cm} (b) \hspace{4.5 cm} (c)
\end{tabular}       
\end{figure}
\begin{figure}[htp]
\begin{tabular}{c}
        \includegraphics[scale=0.3]{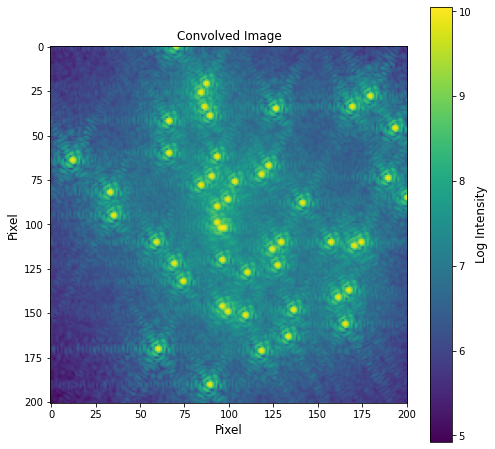}
        \includegraphics[scale=0.3]{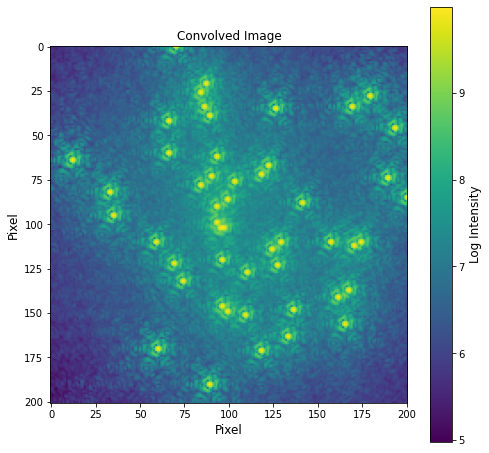}
        \includegraphics[scale=0.3]{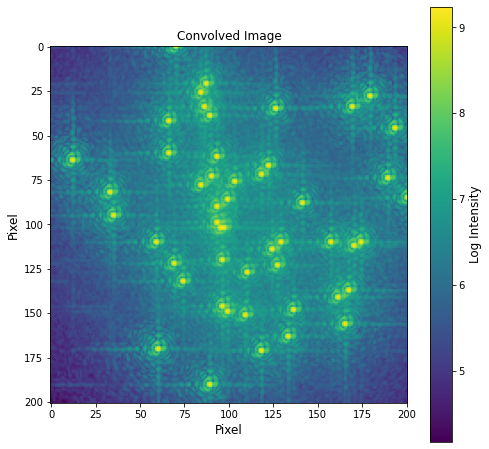}\\
(d) \hspace{4.5 cm} (e) \hspace{4.5 cm} (f)
\end{tabular}
\caption{Convolved images for various apertures for star field of magnitude 0: (a) Circular (b) Elliptical with horizontal major axis (c) Elliptical with vertical major axis (d) Hexagonal (e) JWST (f) HST}
\end{figure}

\subsection{Addition of Photon and Detector Noise}
\label{sect:title}
In the next step, we introduce photon and detector noise to our convolved image. 
Detector noise is present due to various defects in the detector. It can be present even in the absence of any incoming light. Two primary noise sources contributing to detector noise are dark current and read noise. Dark current is a type of noise caused by the spontaneous generation of charge carriers (electrons or holes) within the semiconductor material of an image sensor, even in the absence of external light. It results from thermal energy in the sensor. 
Read noise, on the other hand, arises when incoming photons are converted to a charge in the form of electrons. Any deviation from the ideal value of photon count is called read noise. 
In our case, we consider dark current rate to be equal to \(3\times10^{-3}\) counts/s and read noise to be 2 counts.
\\
Photon noise, also known as shot noise, arises because photons arrive at the detector in a discrete and random manner. In a detector, the number of photons detected per unit time (or per pixel) follows a Poisson distribution. This distribution causes variations in the number of photons detected over time, resulting in shot noise.

\begin{figure}[htp]
\begin{tabular}{c}
        \includegraphics[scale=0.3]{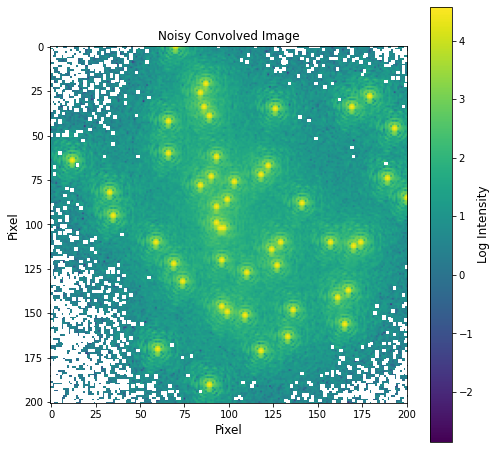}
        \includegraphics[scale=0.3]{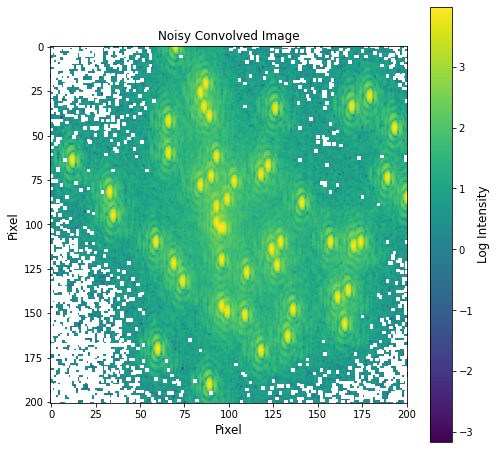}
        \includegraphics[scale=0.3]{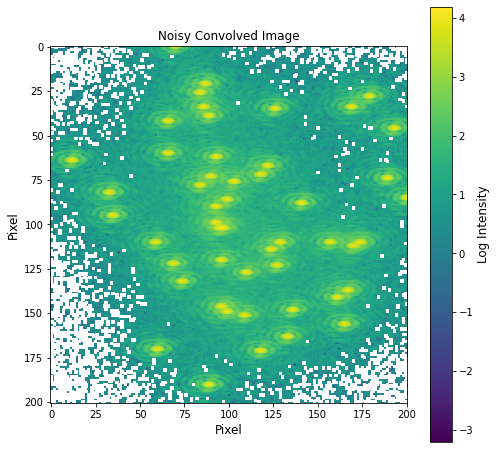}\\
(a) \hspace{4.5 cm} (b) \hspace{4.5 cm} (c)
\end{tabular}       
\end{figure}
\begin{figure}[htp]
\begin{tabular}{c}
        \includegraphics[scale=0.3]{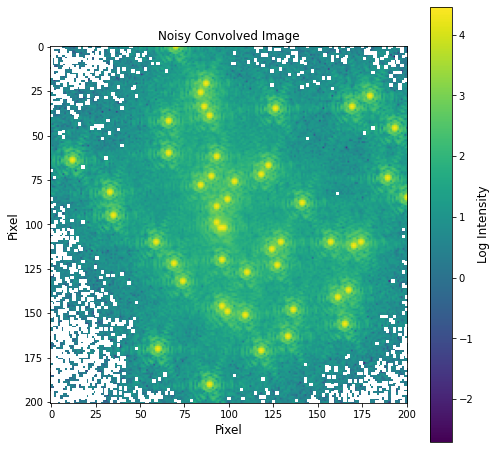}
        \includegraphics[scale=0.3]{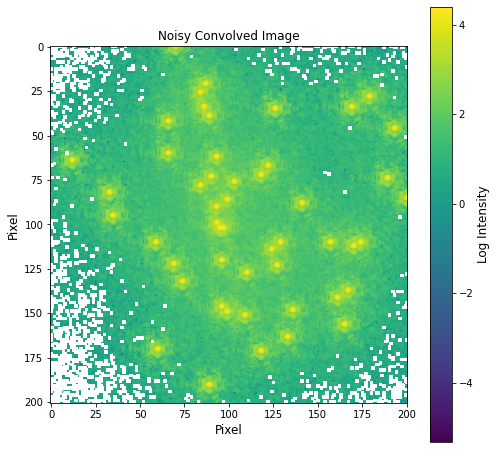}
        \includegraphics[scale=0.3]{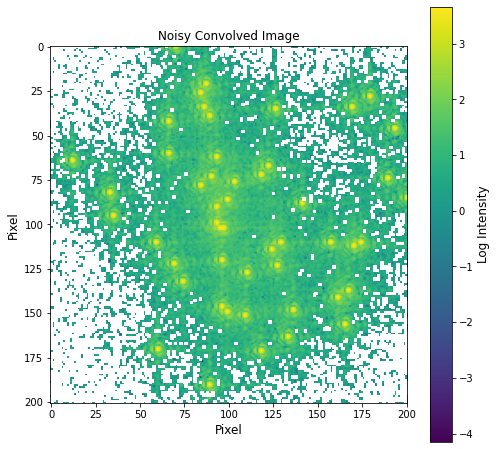}\\
(d) \hspace{4.5 cm} (e) \hspace{4.5 cm} (f)
\end{tabular}
\caption{Noisy Convolved images for various apertures for star field of magnitude 14: (a) Circular (b) Elliptical with horizontal major axis (c) Elliptical with vertical major axis (d) Hexagonal (e) JWST (f) HST}
\end{figure}

\subsection{Deconvolution of Noisy Images}
\label{sect:title}
In the last step of our simulation, we deconvolve the noisy images using the PSF and judge the accuracy of the deconvolution performed. We use the Richardson-Lucy deconvolution algorithm \cite{bioimageanalysisnotebooks} and vary the number of iterations from 10 - 200 to obtain the best possible deconvolved image. We judge deconvolution quality in terms of the number of stars detected, which should be equal to the original value.\\
For magnitudes from 0 to 10, the deconvolution algorithm performs well but it faces challenges in distinguishing stars that are close together, often counting them as one. In such cases, 200 iterations produce a sharper image and detects more stars. \\
At magnitude 12, deconvolution with 10 iterations results in a blurred image of the stars but successfully detects more stars. Increasing the number of iterations further degrades the outcome. Hence, for magnitude 12, we opt for deconvolution with 10 iterations to calculate star positions. \\
Deconvolution at magnitude 14 delivers suboptimal results, with some stars remaining detectable. However, this scenario can yield false positives in certain instances. 

\begin{figure}[htp]
\centering
\begin{tabular}{c}
        \includegraphics[scale=0.3]{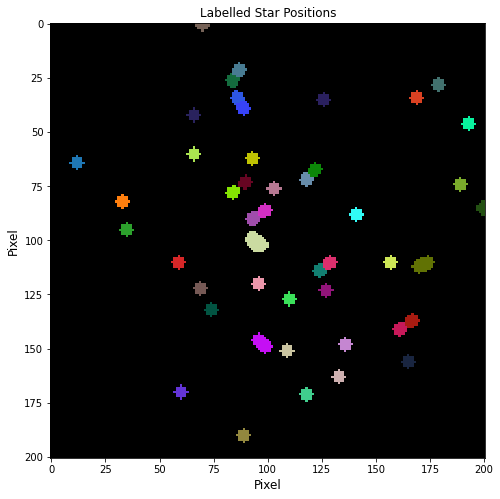}
        \includegraphics[scale=0.33]{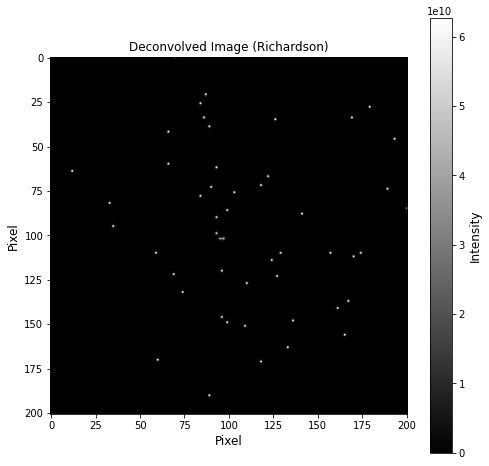}\\
(a) \hspace{4.5 cm} (b)
\end{tabular}  
\caption{Deconvolution of Noisy images: (a) Labelling Star Positions (b) Deconvolved Star Field}
\end{figure}

\section{Image Processing Performance Benchmark}
\label{sect:sections}
In this section, we present the key findings of our study, focusing on two main aspects. Section 3.1 explores the relationship between Signal-to-Noise Ratio (SNR) and stellar magnitude, revealing the trend of the PSF becoming increasingly noisy with higher magnitudes. This analysis provides valuable insights into the behaviour of the PSF under varying brightness conditions, which is crucial for understanding the limitations of astronomical observations. In Section 3.2, we examine the deconvolution accuracy, using two parameters — the average distance between original and calculated star positions and the difference between original and detected number of stars. This analysis sheds light on how well the deconvolution process can recover point sources within the image. We also investigate the influence of different telescope apertures, unraveling the distinctive effects of varied apertures on the observed astronomical scenes. 

\begin{figure}[htp]
        \centering
        \includegraphics[scale=0.3]{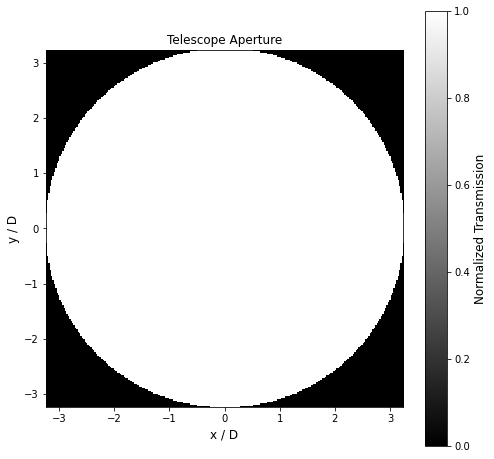}
        \includegraphics[scale=0.3]{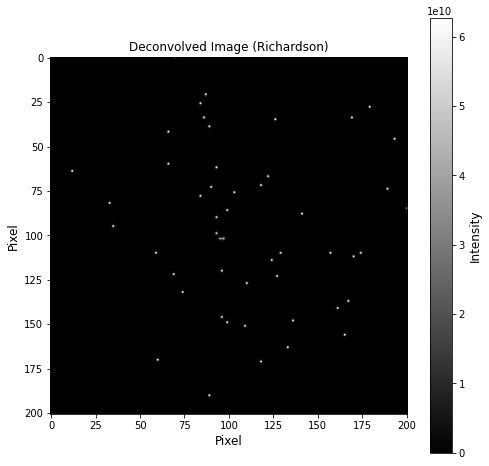}
   \caption{Deconvolved image obtained with Circular Aperture}
\end{figure}

\begin{figure}[htp]
        \centering
        \includegraphics[scale=0.3]{Figures/Hexagonal/hexagonal_ap.png}
        \includegraphics[scale=0.3]{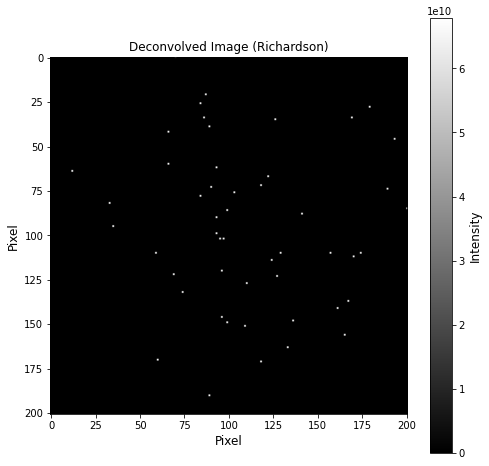}
   \caption{Deconvolved image obtained with Hexagonal Aperture}
\end{figure}

\begin{figure}[htp]
        \centering
        \includegraphics[scale=0.3]{Figures/Elliptical_horizontal_major_axis/elliptical1_ap.png}
        \includegraphics[scale=0.3]{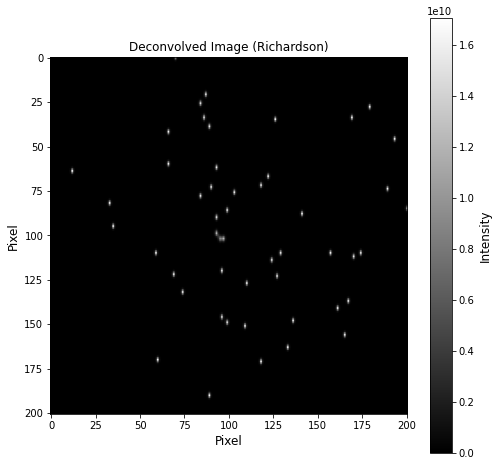}
   \caption{Deconvolved image obtained with Elliptical Aperture having horizontal major axis}
\end{figure}

\begin{figure}[htp]
        \centering
        \includegraphics[scale=0.3]{Figures/Elliptical_vertical_major_axis/elliptical2_ap.png}
        \includegraphics[scale=0.3]{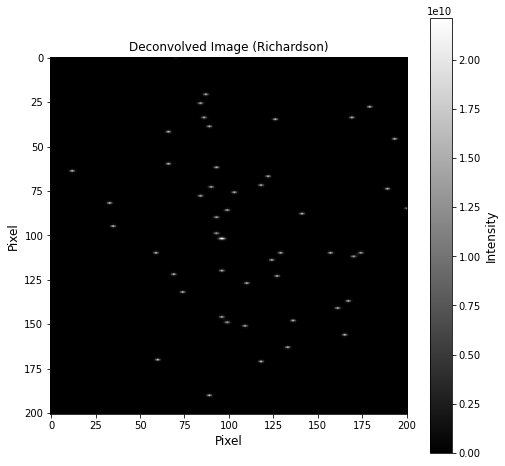}
   \caption{Deconvolved image obtained with Elliptical Aperture having vertical major axis}
\end{figure}

\begin{figure}[htp]
        \centering
        \includegraphics[scale=0.3]{Figures/JWST/jwst_ap.png}
        \includegraphics[scale=0.3]{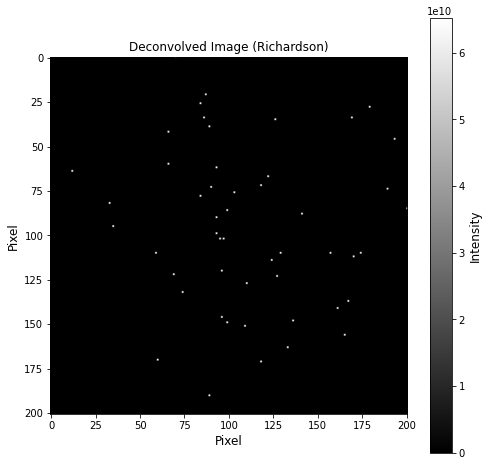}
   \caption{Deconvolved image obtained with JWST Aperture}
\end{figure}

\begin{figure}[htp]
        \centering
        \includegraphics[scale=0.3]{Figures/HST/hst_ap.png}
        \includegraphics[scale=0.3]{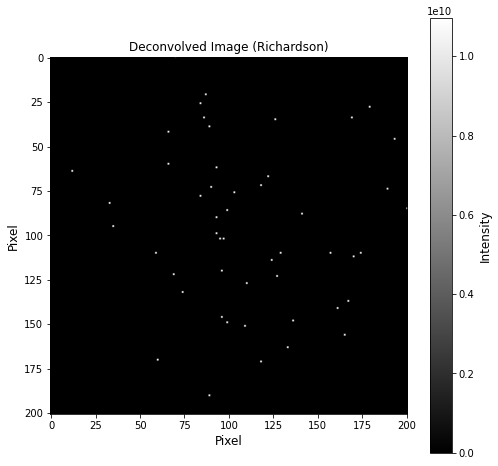}
   \caption{Deconvolved image obtained with HST Aperture}
\end{figure}

\newpage
\subsection{Variation in Signal-to-Noise Ratio with Brightness Magnitude}
\label{sect:title}
We observe that the Signal-to-Noise Ratio (SNR) decreases with decreasing brightness of stars. As brightness decreases, the number of incoming photons decreases, the signal becomes weaker and the noise becomes a larger fraction of the total signal, leading to a lower SNR value. As a result, the PSF also becomes increasingly noisier with decreasing SNR values, as shown in Fig 19.

\begin{figure}[htp]
\centering
        \includegraphics[scale=0.8]{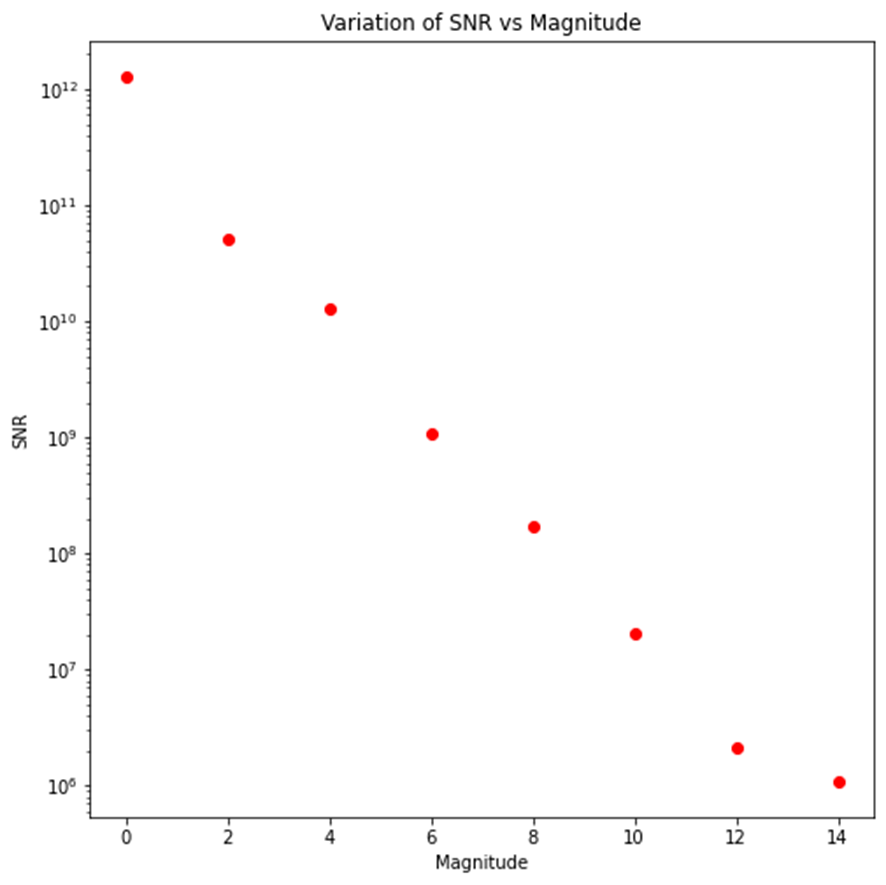}    
\caption{Variation of SNR with Magnitude of Brightness of stars}
\end{figure}

\begin{figure}[htp]
\centering
\begin{tabular}{c}
        \includegraphics[scale=0.22]{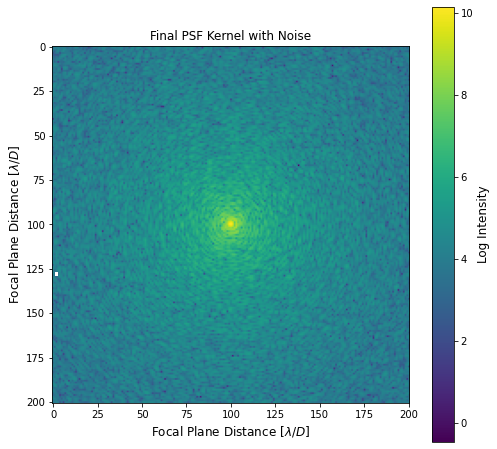}
        \includegraphics[scale=0.22]{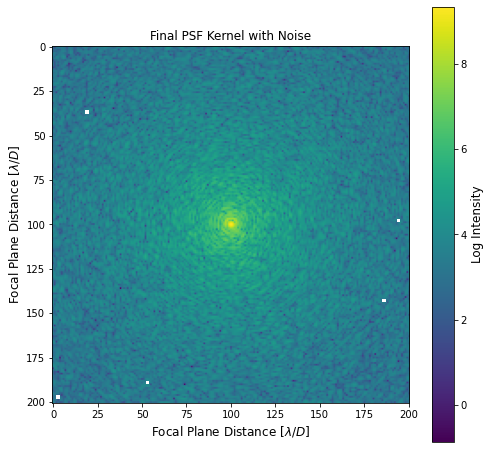}
        \includegraphics[scale=0.22]{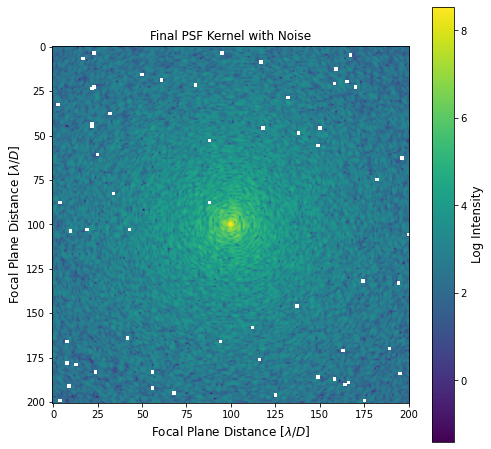}
        \includegraphics[scale=0.22]{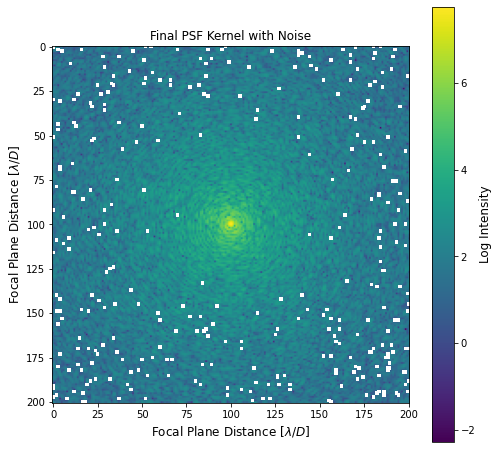}\\
(a) \hspace{3 cm} (b) \hspace{3.5 cm} (c) \hspace{3 cm} (d)
\end{tabular}
\end{figure}

\newpage

\begin{figure}[htp]
\centering
\begin{tabular}{c}
        \includegraphics[scale=0.22]{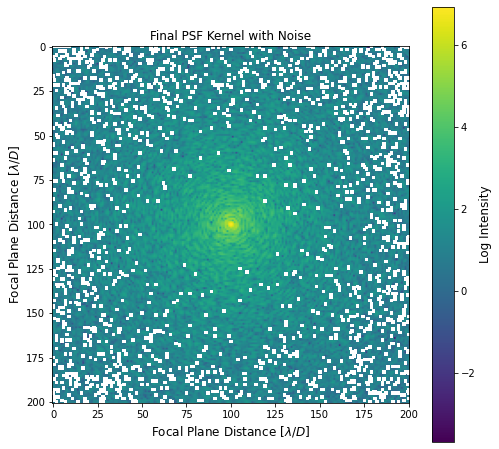}
        \includegraphics[scale=0.22]{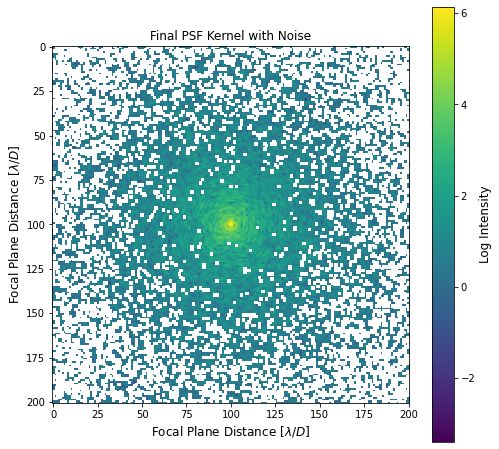}
        \includegraphics[scale=0.22]{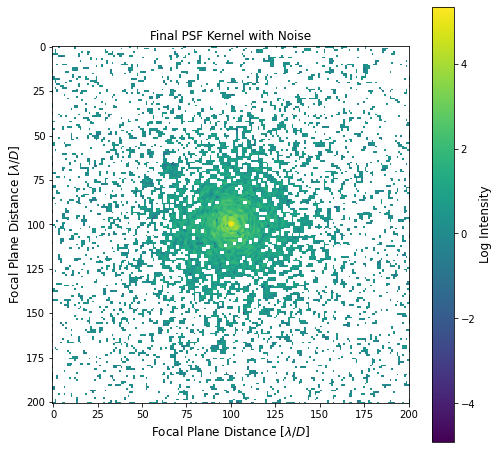}
        \includegraphics[scale=0.22]{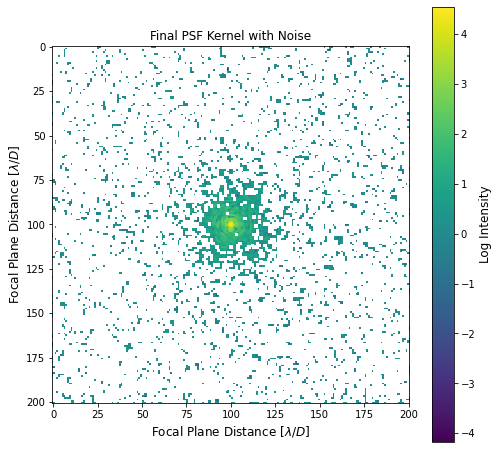}\\
(e) \hspace{3 cm} (f) \hspace{3.5 cm} (g) \hspace{3 cm} (h)
\end{tabular}
\caption{Noisy PSF: (a) Circular (b) Elliptical with horizontal major axis (c) Elliptical with vertical major axis (d) Hexagonal (e) JWST (f) HST}
\end{figure}

When we fix the telescope aperture and vary the brightness of the star field, we observe that photon and detector noise has a more significant impact on faint stars because the noise is a larger fraction of their total signal. This is attributed to the fact that faint stars emit less number of photons so they have lower SNR, as their signal (brightness) is closer to the noise level. Conversely, photon and detector noise has lesser impact on bright stars because they emit large number of photons and have higher SNR, so noise forms a smaller fraction of their total signal.

\begin{figure}[htp]
\centering
\begin{tabular}{c}
        \includegraphics[scale=0.22]{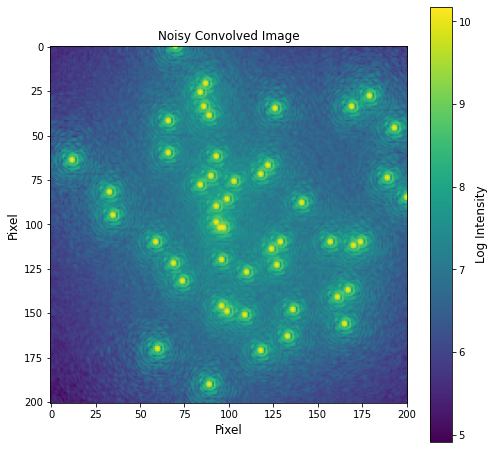}
        \includegraphics[scale=0.22]{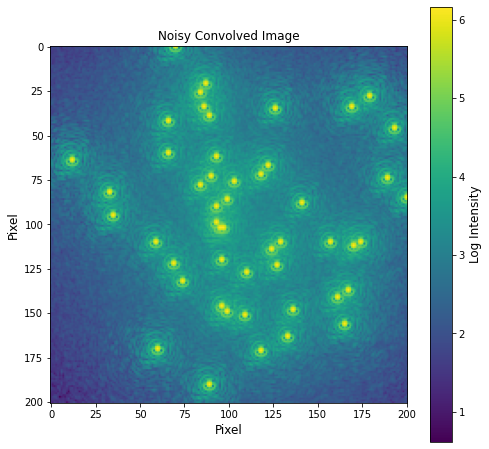}
        \includegraphics[scale=0.22]{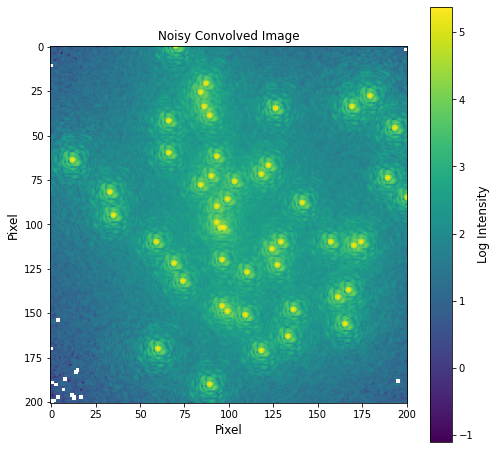}
        \includegraphics[scale=0.22]{Figures/Circular/noisy_conv_14.png}\\
(a) \hspace{3 cm} (b) \hspace{3.5 cm} (c) \hspace{3 cm} (d)
\end{tabular}
\caption{Varying the Brightness of Star Field
and using a Circular Aperture: (a) Magnitude 0 (b) Magnitude 10 (c) Magnitude 12 (d) Magnitude 14}
\end{figure}

\subsection{Assessing Deconvolution Accuracy of Noisy Images}
\label{sect:title}

\subsubsection{Average Distance between Original and Calculated Position of Stars}
\label{sect:title}
After obtaining the deconvolved images, we compare it to the original image, to determine the average distance between the original and calculated positions of stars. We observe that for all telescope apertures, the average distance is relatively stable, having a value close to 0.06 pixels (\(0.7853 \times 10^{-6}\) arcseconds for HST aperture and \(0.2899 \times 10^{-6}\) arcseconds for all other apertures), for most magnitudes of brightness (0 to 12). As brightness decreases, the average distance shoots up suddenly, at magnitude 14. This is because lower SNR introduces uncertainty and affects the precision of calculated positions. Thus, the SNR is one of the factors that affects the accuracy of deconvolution.

\begin{figure}[htp]
\begin{tabular}{c}
        \includegraphics[scale=0.3]{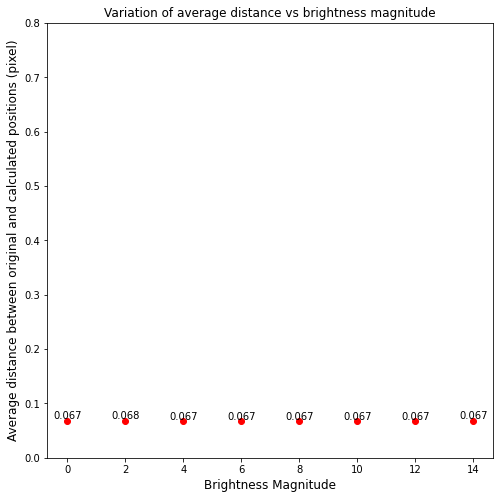}
        \includegraphics[scale=0.3]{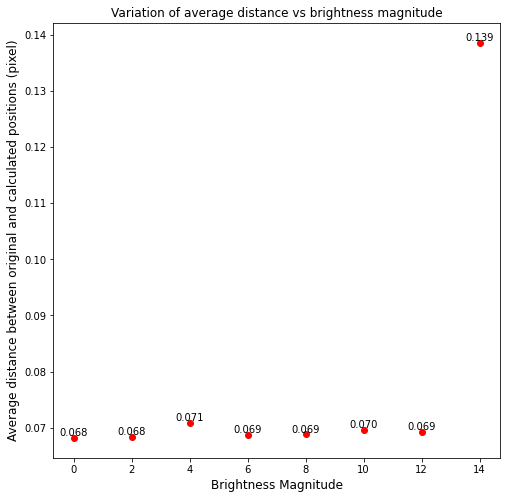}
        \includegraphics[scale=0.3]{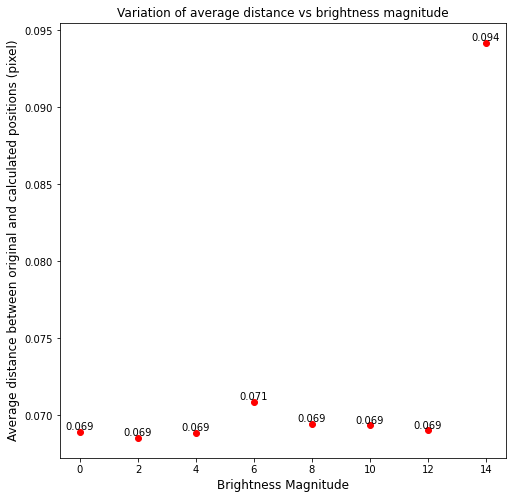}\\
(a) \hspace{4.5 cm} (b) \hspace{4.5 cm} (c)
\end{tabular}       
\end{figure}
\begin{figure}[htp]
\begin{tabular}{c}
        \includegraphics[scale=0.3]{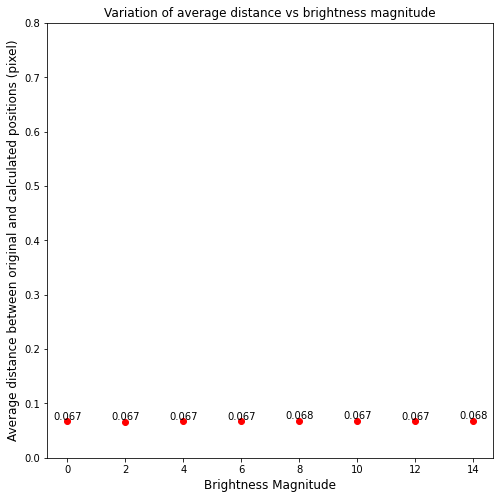}
        \includegraphics[scale=0.3]{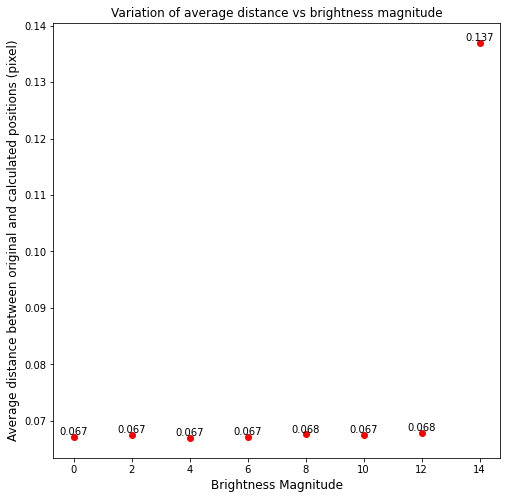}
        \includegraphics[scale=0.3]{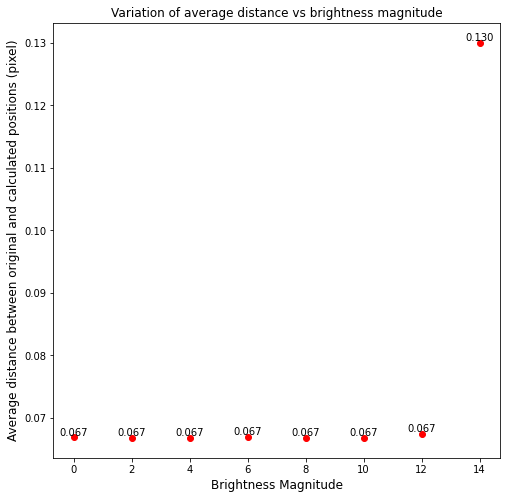}\\
(d) \hspace{4.5 cm} (e) \hspace{4.5 cm} (f)
\end{tabular}
\caption{Average Distance between Original and Calculated Positions for different Magnitudes of Brightness and Apertures: (a) Circular (b) Elliptical with horizontal major axis (c) Elliptical with vertical major axis (d) Hexagonal (e) JWST (f) HST}
\end{figure}

\subsubsection{Difference between Original and Detected Number of Stars}
\label{sect:title}
The count of detected stars shows the reliability of the deconvolution algorithm and effect of SNR and telescope apertures on deconvolution performance. As shown in Figure 22, for all telescope apertures, we observe that the number of stars detected is relatively stable, showing a reduction by 3 stars for most magnitudes of brightness (0 to 12). However, with a decrease in brightness and a consequent decrease in SNR, the number of detected stars drops suddenly, at magnitude 14. Thus, the SNR directly affects the deconvolution performance. The reduction in number of stars detected can also be attributed to the following reasons - a) Some stars are located very close to each other. The deconvolution algorithm is thus unable to distinguish between them and counts them as one. b) Some stars are too near to the edges, making it difficult for the algorithm to detect them.

\begin{figure}[htp]
\begin{tabular}{c}
        \includegraphics[scale=0.3]{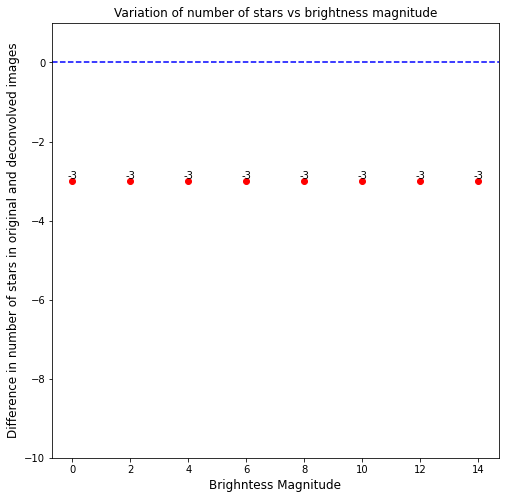}
        \includegraphics[scale=0.3]{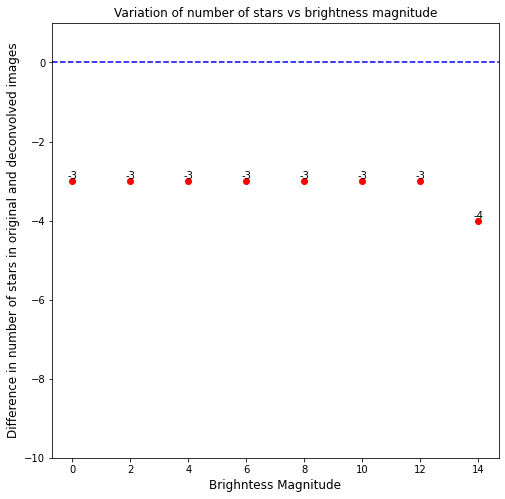}
        \includegraphics[scale=0.3]{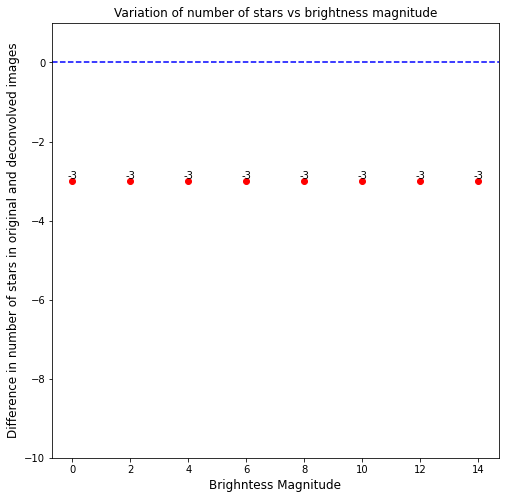}\\
(a) \hspace{4.5 cm} (b) \hspace{4.5 cm} (c)
\end{tabular}       
\end{figure}
\begin{figure}[htp]
\begin{tabular}{c}
        \includegraphics[scale=0.3]{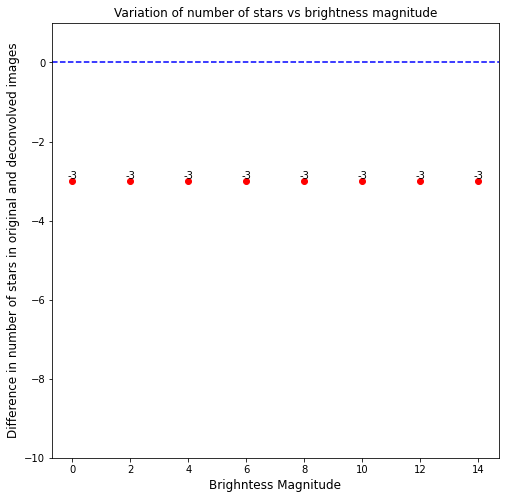}
        \includegraphics[scale=0.3]{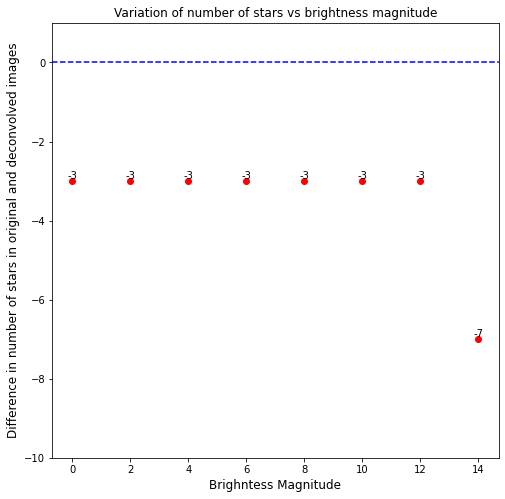}
        \includegraphics[scale=0.3]{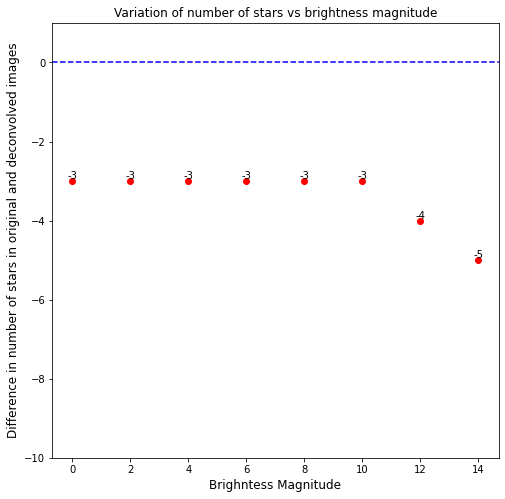}\\
(d) \hspace{4.5 cm} (e) \hspace{4.5 cm} (f)
\end{tabular}
\caption{Difference between Original and Detected Number of Stars for different Magnitudes of Brightness and Apertures: (a) Circular (b) Elliptical with horizontal major axis (c) Elliptical with vertical major axis (d) Hexagonal (e) JWST (f) HST}
\end{figure}

\subsection{Effects of Telescope Apertures}
\label{sect:title}
The choice of telescope aperture plays an important role in influencing both the simulated images and the subsequent deconvolution process. In order to get accurate astronomical images, the Image Processing Pipeline and Deconvolution Algorithms for every aperture has to be optimised in its own way. The shape of the stars in the deconvolved image is also related to the telescope aperture used. In the case of elliptical apertures, we observe an interesting phenomena. When light passes through an aperture, the PSF is determined by the Fourier transform of the aperture function. Hence, in the case of an elliptical aperture with vertical axis, due to the scaling property of Fourier transforms, the resulting PSF will have a major axis rotated in the opposite direction, making it appear like an ellipse with a horizontal major axis. We can also observe this rotation in the deconvolved images where the stars are shaped as elliptical rods, instead of circular dots.\\
Comparing all the apertures, the monolithic circular aperture serves as the best case because it leads to an Airy PSF. For all the brightness magnitudes, the average distance between original and calculated star positions and the difference between original and detected number of stars are minimized for the circular aperture among all the apertures. However, it is still not ideal and has errors which must be optimized computationally. In addition, monolithic circular apertures are difficult to scale for future large-aperture space telescopes due to manufacturing difficulties and limited rocket fairing sizes. Thus, there is always a trade off between the optimal imaging performance and practical considerations related to design and manufacturing.

\section{Conclusions}
We analysed the effects of telescope apertures on the image simulation and deconvolution of a synthetic star field. Using HCIPy and Python programming, we modelled six telescope apertures namely Circular, Hexagonal, Elliptical (with horizontal and vertical major axes respectively), segmented hexagonal (JWST) and obstructed circular (HST). We calculated the PSFs from each aperture, accounted for surface aberrations and used it to convolve with our synthetic star field, varying the brightness magnitude of the stars in the range of 0 to 14. We introduced realistic noise in the images in the form of photon and detector noise. We then performed the deconvolution of these images using the Richardson-Lucy deconvolution algorithm and analysed the accuracy of the deconvolution using metrics like the average distance between original and calculated star positions and the difference between original and detected number of stars. \\
The results obtained show that the choice of telescope aperture plays an important role in influencing both the simulated images and the subsequent deconvolution process. The brightness of the star field (measured as SNR) is also an important factor that should be taken into account. For all apertures, the average distance between original and calculated star positions is approximately 0.06 pixels for brightness magnitudes between 0 and 12. For magnitude 14, the distance shoots up and goes to higher values depending on the aperture type. The absolute difference between original and detected number of stars shows a similar trend - it is equal to 3 for brightness magnitudes lying between 0 and 12 and increases as the magnitude increases (brightness decreases).  These errors in the deconvolution performance allow us to conclude that in order to get accurate astronomical images, the image processing pipeline and deconvolution algorithms for every aperture has to be optimised in its own way. Thus, the choice of telescope aperture and its optimisation pipeline play a very important role.\\
Further study will be required to determine the 'optimal' telescope aperture and its processing pipeline, while also keeping in mind practical considerations related to design and manufacturing.\\
\\
\textbf{Disclosures.} The authors declare no conflicts of interest.\\
\\
\textbf{Data Availability Statement.} Data underlying the results presented in this paper may be obtained from the authors upon reasonable request.

\bibliography{report}   

\begin{thebibliography}{1}

\bibitem{hcipy-docs}
H.~D. Team, {\em HCIPy Documentation}.
\newblock HCIPy  (2023).

\bibitem{inproceedings}
E.~Por, S.~Haffert, V.~Radhakrishnan, {\em et~al.}, ``High contrast imaging for python (hcipy): an open-source adaptive optics and coronagraph simulator,'' 152  (2018).

\bibitem{martiniusefuldata}
P.~Martini, {\em Useful Data for Astronomers}.
\newblock \linkable{astronomy.ohio-state.edu/martini.10/usefuldata}.

\bibitem{bioimageanalysisnotebooks}
S.~H{\"a}sleinh{\"u}pf, {\em Bio Image Analysis Notebooks}  (2023).
\newblock \linkable{haesleinhuepf.github.io/BioImageAnalysisNotebooks}.

\end{thebibliography}
\bibliographystyle{spiejour}   

\end{spacing}
\end{document}